\documentclass[12pt]{iopart}
\usepackage{iopams}
\usepackage{verbatim}
\usepackage[dvips]{epsfig}

\def\##1{{\bf #1}}
\def\=#1{\underline{\underline #1}}

\def\.{\mbox{ \tiny{$^\bullet$} }}

\def\eps{\varepsilon}
\def\epso{{\eps}_o}

\def\ux{\#u_x}
\def\uy{\#u_y}
\def\uz{\#u_z}
\def\zz{\uz\uz}

\def\muo{\mu_o}
\def\lambdao{\lambda_o}

\def\ko{k_o}
\def\co{c_o}

\def\les{\left[}
\def\ris{\right]}

\def\epsa{\eps_\parallel}
\def\epsb{\eps_\perp}
\def\mua{\mu_\parallel}
\def\mub{\mu_\perp}
\def\xia{\xi_\parallel}
\def\xib{\xi_\perp}

\begin{document}

\title[Spectral shifts due to isotropic chirality
]{Spectral shifts in the properties of a periodic multilayered stack due to isotropic chiral layers}

\author{S. Anantha Ramakrishna$^{1,2}$\footnote{E-mail:anantha@iisermohali.ac.in}  and Akhlesh Lakhtakia$^{1,3}$\footnote{Corresponding author;
e-mail:akhlesh@psu.edu}}
\address{$^{1}$Department of Physics, Indian Institute of Technology Kanpur, Kanpur 208016, India}
\address{$^{2}$Department of Physics, Indian Institute of Science Education and Research Mohali, Transit Campus, MGSIPA Complex, Sector-26, Chandigarh  160019, India}
\address{$^{3}$NanoMM---Nanoengineered Metamaterials Group, Department of Engineering Science and Mechanics, Pennsylvania State University, University Park, PA 16802, USA}

\begin{abstract} Investigating the canonical problem of a periodic multilayered stack containing
isotropic  chiral layers, we homogenized it as a uniaxial bianisotropic medium and derived its effective
constitutive parameters. The stack  shows a resonant behavior, when its unit cell
consists of a metallic layer and an isotropic
chiral layer. The presence of isotropic chirality
can result in small shifts of the resonance frequency for reasonably large
values of the chirality parameter, implying that the sign of an effective
permittivity can  be switched. Such  spectral shifts in the
dielectric properties can be potentially useful for spectroscopic
purposes.

\vspace{0.5cm}

\end{abstract}

 \maketitle

\section{Introduction}\label{intro}

Metamaterials that refract negatively \cite{Rama-rpp,Wood} have been
demonstrated across the electromagnetic spectrum from microwave to optical
frequencies. A metamaterial's response characteristics and frequency range of operation are
principally determined by the structures of its constituent units. There is,
however, a need to be able to dynamically tune the response of a given
metamaterial once it has been fabricated. Kerr
nonlinearity~\cite{obrien_PRB} and the electro-optic
effect~\cite{padilla_Nature2008} have already been suggested in this
regard. These effects, however, are primarily dielectric effects. A related question
arises: can isotropic chirality can be utilized to manipulate the
properties of a metamaterial wherein an isotropic chiral material has been embedded?

The rotation of the vibration ellipse of light after passage through an
isotropic chiral material has been known for about two centuries \cite{Lspie}. In the frequency domain,
this kind of material is described by three constitutive parameters: relative permittivity,
relative permeability, and chirality parameter. At least one of the three constitutive parameters
has to be different from its free-space counterpart in order that the material be distinguishable
from free space (i.e., vacuum). For an isotropic chiral material to be negatively refracting~\cite{Lemc}-\cite{JPen}, a minimum
of two constitutive parameters must be different from their free-space counterparts \cite{LMpre,Mackay-chi}.
Thus, the inclusion of isotropic chirality enlarges the potential for fabricating  isotropic, homogeneous, and negatively refracting materials that are actually useful.

For an isotropic chiral material to be negatively refracting, the chirality parameter must be sufficiently large in magnitude
compared to the product of the relative permittivity and the relative permeability, the underlying assumption being that the
material is nondissipative at the frequency of interest \cite{JPen}.  If dissipation is present, the condition for negative refraction
becomes more complicated, but, in essence, the magnitude of
the chirality parameter is still required to be large \cite{Mackay-chi}. In nature, large chirality parameters (in the optical regime) are not known to
exist \cite{Bohren}, and very large chirality parameters for artificial materials (in the microwave
regime) have not been reported yet \cite{VRV,CW,Gomez}. Therefore, for
isotropic  chirality to be effective in delivering the attribute of negative refraction,
the magnitude of either the relative permittivity or the relative permeability must be close to zero \cite{JPen}.
For dynamic control of a metamaterial using isotropic chirality, the fact that the chirality parameter can dominate 
relative permittivity and/or relative permeabilty
becomes particularly relevant in the frequency ranges where either of them goes through a zero.

In this communication, we investigate whether the isotropic chirality of an embedded
medium can be effectively used to strongly influence the spectral properties of a
metamaterial by choosing the canonical problem of a periodic multilayered stack that can be homogenized.
Provided the periodic multilayered stack consists of metallic layers and isotropic chiral layers,
we show that the latter type of layers can even switch the sign of
the real part of the effective permittivity of the stack. The
associated frequency shifts are of the order of an {\aa}ngstr\"om in the optical regime, which are 
easily measurable~\cite{OOHR4000} and could be useful for spectroscopic purposes.

The plan of this communication is as follows. Section~\ref{form} describes two different formulations
to homogenize a
periodic  multilayered stack whose unit cell contains two layers both of which can be made of
isotropic chiral materials. The homogenized stack is a uniaxial bianisotropic medium (UBM)~\cite{Weigl94}.
Numerical results on spectroscopic shifts are discussed in Section~\ref{res}, followed by concluding remarks
in Section~\ref{conc}.
An $\exp(-i\omega t)$ time-dependence is implicit, with $\omega$ denoting the angular frequency. The free-space wavenumber, the free-space wavelength, and the speed of light in free space are denoted by $\ko=\omega\sqrt{\epso\muo}$, $\lambdao=2\pi/\ko$, and $\co=1/\sqrt{\muo\epso}$, respectively, with $\muo$ and $\epso$ being  the permeability and permittivity of free space. Vectors are in boldface, dyadics   underlined twice; column vectors are in boldface and enclosed within square brackets, whereas matrixes are underlined twice and similarly bracketed. Cartesian unit vectors are identified as $\ux$, $\uy$ and $\uz$. The dyadics employed in the following sections can be treated as 3$\times$3 matrixes \cite{Chen}.

\section{Theory}\label{form}
\begin{figure}
\begin{center}
 \begin{tabular}{cc}
\includegraphics[height=8cm]{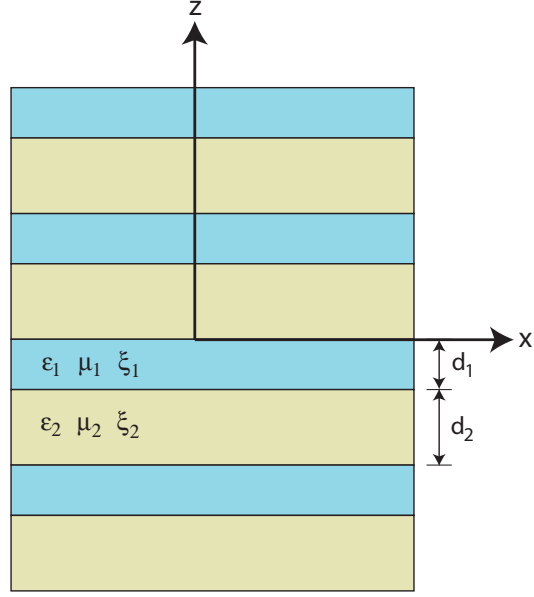}
\end{tabular}
 \end{center}
\caption{ \label{Fig1} Schematic of the periodic multilayered stack of interest.}
\end{figure}

Let us consider a periodic multilayered stack whose unit cell is made of two layers of dissimilar
materials labeled $1$ and $2$, as shown in Fig.~\ref{Fig1}. Both materials are isotropic chiral, with their Tellegen constitutive
relations written in the frequency domain as~\cite{Lspie}-\cite{Mackay-chi}
\begin{equation}
\left.\begin{array}{l}
\#D=\epso\eps_n\#E + i \co^{-1}\xi_n\#H\\
\#B=\muo\mu_n\#H - i \co^{-1}\xi_n\#E
\end{array}
\right\}\,,\quad n= 1,2\,,
\label{conrel-n}
\end{equation}
where $\eps_n$ is the relative permittivity, $\mu_n$ is the relative permeability, and
$\xi_n$ is the chirality parameter. The thickness of the  layer made of the $n$-th material is
denoted by $d_n$, so that
\begin{equation}
f_n= d_n/(d_1+d_2)\,,\quad n= 1,2\,,
\end{equation}
is the volume fraction of the $n$-th material in the stack. 

Let the $z$ axis of a Cartesian coordinate system be oriented normal
to the layers.
Provided that both
layers in the unit cell are electrically thin, the periodic stack can be homogenized as
a uniaxial bianisotropic medium (UBM) whose distinguished axis is the $z$ axis \cite{Weigl94}. The
frequency-domain constitutive relations of the UBM are given as
\begin{equation}
\label{constitutive_relations}
\left.\begin{array}{l}
\#D=\epso\les\epsb\,\=I+(\epsa-\epsb)\zz\ris\.\#E\\\qquad\qquad+ i\co^{-1}\les\xib\,\=I+(\xia-\xib)\zz\ris\.\#H\\[5pt]
\#B=\muo\les\mub\,\=I+(\mua-\mub)\zz\ris\.\#H \\\qquad\qquad- i\co^{-1}\les\xib\,\=I+(\xia-\xib)\zz\ris\.\#E
\end{array}\right\}\,,
\end{equation}
where $\=I$ is the identity dyadic, and there are six \emph{effective} constitutive parameters: $\epsa$, $\epsb$,
$\mua$, $\mub$, $\xia$, and $\xib$. In the following subsections, we provide two different ways
to determine these six scalars in terms of the constitutive parameters and the volume fractions
of the two constituent materials in the unit cell.

\subsection{4$\times$4-Matrix method}\label{1stmethod}
Without loss of generality,
wave propagation in the multilayered stack can be handled by using the spatial Fourier transform as
\begin{equation}
\left.\begin{array}{l}
\#E(x,y,z)=\int_{-\infty}^\infty\,\#e(z,\kappa) \exp\left(i \kappa x\right)\, d\kappa\\[5pt]
\#H(x,y,z)=\int_{-\infty}^\infty\,\#h(z,\kappa) \exp\left(i \kappa x\right)\,d\kappa
\end{array}
\right\}\,, 
\end{equation}
whereby the fields are assumed not to vary along the $y$ axis. The spatial frequency
along the $x$ axis is denoted by $\kappa$. Substitution of this representation along with the
constitutive relations (\ref{conrel-n}) in the two Maxwell curl equations leads to the 4$\times$4-matrix
ordinary differential equation \cite{L-Brew}
\begin{equation}
\frac{d}{dz}\les\#F(z,\kappa)\ris=i\les\=P^{(n)}(\kappa)\ris\.\les\#F(z,\kappa)\ris
\end{equation}
in any layer made of the $n$-th material. Here,
the column vector
\begin{equation}
\les\#F(z,\kappa)\ris= \les\begin{array}{c} e_x(z,\kappa)\\ e_y(z,\kappa)\\ h_x(z,\kappa) \\ h_y(z,\kappa)\end{array}\ris
\end{equation}
represents components of the electric and magnetic fields that are tangential to the bimaterial
interfaces in the multilayered stack, whereas the 4$\times$4 matrix
\begin{eqnarray}
\nonumber
&&\les\=P^{(n)}(\kappa)\ris=\omega\left[\begin{array}{cccc}
0 &-i\xi_n/\co & 0 & \muo\mu_n\\
i\xi_n/\co & 0 &-\muo\mu_n & 0\\
0 &-\epso\eps_n& 0 & -i\xi_n/\co\\
\epso\eps_n & 0& i\xi_n/\co & 0
\end{array}\ris
\\[5pt]
&&\qquad
+\frac{\kappa^2}{\omega\left(\eps_n\mu_n-\xi_n^2\right)}\,
\left[\begin{array}{cccc}
0 & -i\xi_n\co & 0 &-\mu_n/\epso\\
0 & 0 & 0 & 0\\
0 &\eps_n/\muo&0 &-i\xi_n\co\\
0 & 0 & 0 & 0
\end{array}\ris\,,\quad n= 1,2\,,
\end{eqnarray}
contains all three constitutive parameters, the angular frequency, and the parameter $\kappa$ denoting propagation along the $x$ axis.

A similar exercise for wave propagation in the homogenized medium (i.e., the equivalent UBM) yields the
4$\times$4-matrix
ordinary differential equation \cite{L-Brew}
\begin{equation}
\frac{d}{dz}\les\#F(z,\kappa)\ris=i\les\=P^{eqvt}(\kappa)\ris\.\les\#F(z,\kappa)\ris\,,
\end{equation}
where
\begin{eqnarray}
\nonumber
&&\les\=P^{eqvt}(\kappa)\ris=\omega\left[\begin{array}{cccc}
0 &-i\xib/\co & 0 & \muo\mub\\
i\xib/\co & 0 &-\muo\mub & 0\\
0 &-\epso\epsb& 0 & -i\xib/\co\\
\epso\epsb & 0& i\xib/\co & 0
\end{array}\ris
\\[5pt]
&&\qquad
+\frac{\kappa^2}{\omega\left(\epsa\mua-\xia^2\right)}\,
\left[\begin{array}{cccc}
0 & -i\xia\co & 0 &-\mua/\epso\\
0 & 0 & 0 & 0\\
0 &\epsa/\muo&0 &-i\xia\co\\
0 & 0 & 0 & 0
\end{array}\ris\,.
\label{P-ubm}
\end{eqnarray}

As both layers in the unit cell are electrically thin,
we can invoke the long-wavelength approximation \cite{LK} to set
\begin{equation}
\les\=P^{eqvt}(\kappa)\ris=f_1\les\=P^{(1)}(\kappa)\ris+f_2\les\=P^{(2)}(\kappa)\ris\,.
\end{equation}
With
\begin{equation}
\Delta = (f_1\varepsilon_2 + f_2 \varepsilon_1)(f_1\mu_2 + f_2\mu_1) - (f_1 \xi_2+f_2 \xi_1)^2\,,
\end{equation}
this procedure yields the following expressions for the effective constitutive parameters of the
stack:
\begin{eqnarray}
\label{epsb}
&&\epsb=f_1\eps_1+f_2\eps_2\,,
\\[5pt]
\label{epsa}
&&\epsa=\Delta^{-1}\les{f_1\eps_1(\eps_2\mu_2-\xi_2^2)+f_2\eps_2(\eps_1\mu_1-\xi_1^2)}\ris\,,
\\[5pt]
\label{mub}
&&\mub=f_1\mu_1+f_2\mu_2\,,
\\[5pt]
\label{mua}
&&\mua=\Delta^{-1}\les{f_1\mu_1(\eps_2\mu_2-\xi_2^2)+f_2\mu_2(\eps_1\mu_1-\xi_1^2)}\ris\,,
\\[5pt]
\label{xib}
&&\xib=f_1\xi_1+f_2\xi_2\,,
\\[5pt]
\label{xia}
&&\xia=\Delta^{-1}\les{f_1\xi_1(\eps_2\mu_2-\xi_2^2)+f_2\xi_2(\eps_1\mu_1-\xi_1^2)}\ris\,.
\end{eqnarray}
Thus, whereas the mixing of the constitutive parameters of the two layers of the unit cell
is trivially simple
 in the transversely isotropic components ($\epsb$, $\mub$, $\xib$) of
 the constitutive dyadics of the equivalent UBM, the mixing is far richer in the
 axial components ($\epsa$, $\mua$, $\xia$) of those dyadics. Our focus is on these axial components.
 
 \subsection{Method of boundary conditions}\label{2ndmethod}
 
A critical question that arises in any homogenization procedure is, whether the procedure 
preserves the boundary conditions across interfaces that are imposed by the Maxwell
equations on the electromagnetic fields. In this 
subsection, we obtain the effective constitutive dyadics of the periodic multilayered stack by considering the boundary 
conditions on the electromagnetic fields. The obtained parameters are identical to the 
ones obtained by the procedure of Sec.~\ref{1stmethod}, which is a matter of consistency.

In the  limit of very small layer thickness (compared to the wavelength of light), the
electromagnetic fields can be assumed to be reasonably uniform across a unit cell of the periodic stack. But 
they have to satisfy the appropriate  boundary conditions across the bimaterial interfaces.   Suppose the fields
in layer $n$, $n\in\les1,2\ris$, of the unit cell are denoted by $\#E_n$,  $\#H_n$, $\#D_n$, and  $\#B_n$.
Let us first consider the continuity of the tangential components (i.e., oriented parallel to the
$xy$ plane) of the $\mathbf{E}$ 
and $\mathbf{H}$ fields:
\begin{eqnarray}
	\mathbf{E}_{1tan} &= \mathbf{E}_{2tan} &= \langle\mathbf{E}_{tan}\rangle, \\
	\mathbf{H}_{1tan} &= \mathbf{H}_{2tan} &= \langle\mathbf{H}_{tan}\rangle, 
\end{eqnarray}
where $\langle\mathbf{E}_{tan}\rangle$ and $\langle\mathbf{H}_{tan}\rangle$ are the volume-averaged fields. Now considering the volume-averaged fields $\langle\mathbf{D}_{tan}\rangle$ and 
$\langle\mathbf{B}_{tan}\rangle$ defined as
\begin{eqnarray}
	\langle\mathbf{D}_{tan}\rangle &= f_1 \mathbf{D}_{1{tan}} + f_2 \mathbf{D}_{2{tan}} &= \eps_0 \epsb \langle\mathbf{E}_{tan}\rangle + i  \co^{-1}\xib \langle\mathbf{H}_{tan}\rangle, \\ 
	\langle\mathbf{B}_{tan}\rangle &= f_1 \mathbf{B}_{1{tan}} + f_2 \mathbf{B}_{2{tan}} &= \mu_0 \mub \langle\mathbf{H}_{tan}\rangle - i  \co^{-1}\xib\langle\mathbf{E}_{tan}\rangle, 
\end{eqnarray}
we obtain
\begin{eqnarray}
     	\epsb &=& f_1 \eps_1 + f_2\eps_2,\\
	\mub &=& f_1 \mu_1 + f_2\mu_2, \\
	\xib &=& f_1 \xi_1 + f_2\xi_2
\end{eqnarray}
for the equivalent UBM. These results are identical to those given by Eqs.~(\ref{epsb}), (\ref{mub}), and (\ref{xib}), respectively.

Next, let us consider the continuity of the electromagnetic  field components normal to the
bimaterial interfaces (i.e., oriented along the $z$ axis) as follows:
\begin{eqnarray}
	\langle\mathbf{D}_{norm}\rangle = \mathbf{D}_{1{norm}} = \mathbf{D}_{2{norm}}, \\
	\langle\mathbf{B}_{norm}\rangle = \mathbf{B}_{1{norm}} = \mathbf{B}_{2{norm}}, \\
	\langle\mathbf{E}_{norm}\rangle = f_1 \mathbf{E}_{1{norm}} + f_2 \mathbf{E}_{2{norm}}, \\ 
	\langle\mathbf{H}_{norm}\rangle = f_1 \mathbf{H}_{1{norm}} + f_2 \mathbf{H}_{2{norm}}.
\end{eqnarray}
 Using the constitutive relations  (\ref{constitutive_relations}) of the homogenized material, we obtain the following 
 three relations from the foregoing equations:
 \begin{eqnarray}
\frac{\varepsilon_\parallel}{\varepsilon_\parallel\mu_\parallel - \xi_\parallel^2} &=&
f_1  \frac{\varepsilon_1}{\varepsilon_1\mu_1 - \xi_1^2}  + f_2  \frac{\varepsilon_2}{\varepsilon_2\mu_2 - \xi_2^2}\,, \\
\frac{\mu_\parallel}{\varepsilon_\parallel\mu_\parallel - \xi_\parallel^2} &=&
f_1  \frac{\mu_1}{\varepsilon_1\mu_1 - \xi_1^2}  + f_2  \frac{\mu_2}{\varepsilon_2\mu_2 - \xi_2^2}\,, \\
\frac{\xi_\parallel}{\varepsilon_\parallel\mu_\parallel - \xi_\parallel^2} &=&
f_1  \frac{\xi_1}{\varepsilon_1\mu_1 - \xi_1^2}  + f_2  \frac{\xi_2}{\varepsilon_2\mu_2 - \xi_2^2}\,.
\end{eqnarray}
The solutions of these three equations are given by Eqs.~(\ref{epsa}), (\ref{mua}), and (\ref{xia}).
Let us reiterate that, as
all expressions derived in this section involve only the
volume fractions and not the individual layer thicknesses, they are valid only in the 
limit of small layer thicknesses.

The quantities $\epsa$, $\mua$, and $\xia$ satisfy the relation
\begin{equation}
\label{rel}
{\rm Det}
\left[\begin{array}{ccc}
\varepsilon_1   &\varepsilon_2 & \varepsilon_\parallel\\[5pt]
\mu_1   &\mu_2 & \mu_\parallel\\[5pt]
\xi_1   &\xi_2 & \xi_\parallel
\end{array}\right]=0\,.
\end{equation}
This relation implies that both homogenization procedures yield only two of the
three quantities $\epsa$, $\mua$, and $\xia$ independently; the third can be obtained from
Eq.~(\ref{rel}). This feature of homogenization has been remarked upon earlier for dispersions of
electrically small,
isotropic chiral spheres in an isotropic achiral host material
\cite{LVV-JMR}.

\section{Numerical Results and Discussion}\label{res}

In order to prove the premise of this paper, let us set material $1$ to be achiral, i.e., $\xi_1=0$.
Furthermore, without loss of generality, let both materials $1$ and $2$ be nonmagnetic in the
Tellegen representation: $\mu_1=\mu_2=1$.\footnote{An isotropic chiral material may be nonmagnetic
in the Tellegen representation, but not in the Drude-Born-Fedorov   representation,
and vice versa \cite{LVV-JMR}.} In that case, $\mub=1$ whereas
\begin{equation}
\mua= \frac{(f_1\eps_2+f_2\eps_1) - f_1\xi_2^2}{(f_1\eps_2+f_2\eps_1) - f_1^2\xi_2^2}\,.
\end{equation}
Now, we can explore the effect of $\xi_2$ on the sign of ${\rm Re}\left(\epsa\right)$.

Suppose further that material $1$ is silver, so that \cite{HN2007}
\begin{equation}
\eps_1(\lambda_o)=5.7+0.4i - \left(\frac{9q}{2\pi\co\hbar}\right)^2\lambda_o^2\,.
\end{equation}
where $q=1.6022\times10^{-19}$~C is the charge of an electron and $\hbar=1.0546\times10^{-34}$~J~s is the reduced Planck constant.
This expression is valid for $\lambda_o\in\les300,900\ris$~nm.  It has been used, for example, by Pendry \cite{Pen-PL}
and also fits the data of Stahrenberg {\it et al.} \cite{SHWERL} in the chosen spectral regime.
For $\eps_2$, we choose the Lorentz model
\begin{equation}
\eps_2(\lambda_o) = 2+
\frac{C_2}{    1-\left(\frac{\lambda_2}{\lambda_o}\right)^2
-\frac{i\gamma_2\lambda_2}{2\pi\co}\,\left(\frac{\lambda_2}{\lambda_o}\right)   }
\,,
\end{equation}
 wherein $C_2$, $\lambda_2$, and $\gamma_2$ are constants. The values of these parameters chosen for
 the following studies are typical of solid materials; in addition, they keep the magnitudes of $\eps_2$
 from becoming unphysically large in the resonance regime.
 
\begin{figure}
\begin{center}
 \begin{tabular}{cc}
\includegraphics[height=8.5cm]{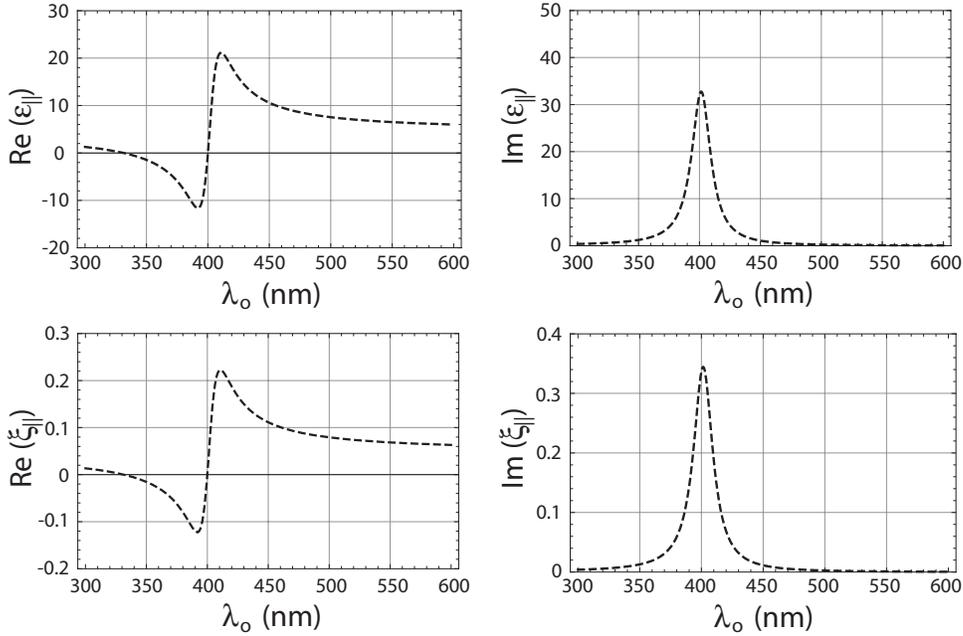}
\end{tabular}
 \end{center}
\caption{ \label{Fig2}  Real and imaginary parts of $\epsa$,
and  real and  imaginary parts of $\xia$, as  functions of free-space wavelength,
when $f_1=0.58$, $C_2=0$, and  $\xi_2=0.05$. }
\end{figure}

 Figure~\ref{Fig2} contains plots of $\epsa$ and $\xia$ with respect to free-space wavelength,
 when material $2$ is nondispersive and nondissipative, and $\xi_2=0.05$. The resonant frequency evident in these plots
 is due to the condition $\Delta = 0$; hence, it is
 fixed by the volume fraction of silver ($f_1=0.58$ for this figure)  and is affected by the dispersive
 properties of silver. The width of the resonance is proportional to the volume fraction of
 the dissipative material (i.e., silver). For physically realistic values of $\xi_2$ ($\lesssim 10^{-3}$), the 
 resonance condition $\Delta=0$ can be achieved only if the ratio ${\rm Re}(\eps_2)/{\rm Re}(\eps_1) < 0$.
 
 The resonance condition enhances $\xia$. This becomes clear by noting that the maximum value of
 $\vert\xia\vert$ is
 many times larger than that of the chirality parameter $\xi_2$ of the isotropic chiral constituent of the
 multilayered stack. In contrast, Eq.~(\ref{xib}) indicates that $\xib$ is diluted in relation to $\xi_2$ by the volume fraction of
material $2$.

The resonance condition also enhances ${\rm Im}(\epsa)$ far above the imaginary part of the
relative permittivity of silver. This is deleterious to wave propagation inside the multilayered stack.
Still, we cannot help remarking that frequency regimes exist (e.g., $\lambdao\approx
500$~nm) wherein  $\xia$ is significantly enhanced whereas  ${\rm Im}(\epsa)$ is significantly reduced
in relation to ${\rm Im}(\eps_1)$.

Let us also note that the homogenized medium, a UBM, has the properties of a chiroplasma except for a gyrotropic
term, because such a term is missing in both constituent materials \cite{WL98}. It also presents an example of a
homogeneous medium
that displays chirality along with a negative real permittivity at optical frequencies, which has not been reported
earlier, to our knowledge.

\begin{figure}
\begin{center}
 \begin{tabular}{cc}
\includegraphics[height=4cm]{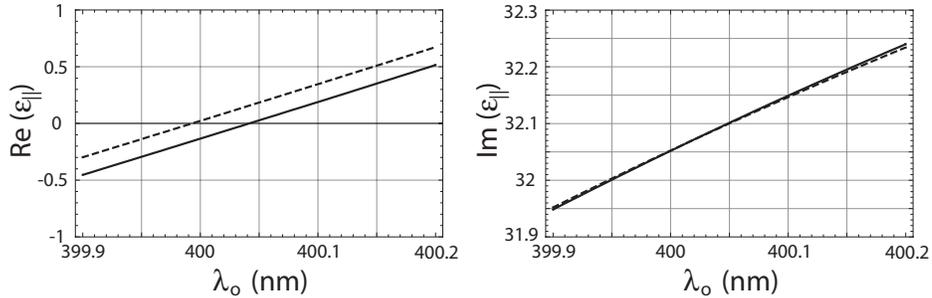}
\end{tabular}
 \end{center}
\caption{ \label{Fig3} Real and  imaginary parts of $\epsa$ as  functions of free-space wavelength,
when $f_1=0.58$ and $C_2=0$; $\xi_2=0$ (solid lines) and $\xi_2=0.05$ (dashed lines).}
\end{figure}

 Figure~\ref{Fig3} presents a highly magnified view of $\epsa$ in the resonance region ($\lambdao\approx
 400$~nm). The constitutive properties and geometric parameters are the same as in the previous figure, 
 except that data is shown for both $\xi_2=0$ and $\xi_2=0.05$. Clearly, the zero-crossing of ${\rm Re}(\epsa)$
 blueshifts by about {$0.5$~\AA}, when $\xi_2$ increases from $0$ to $0.05$, furthermore, the same
 blueshift would occur even if $\xi_2$ were to be replaced by $-\xi_2$.  This spectroscopic shift is measurable~\cite{OOHR4000}.

 We have not shown spectral plots of $\mua$, because its magnitude is very close to unity. It does, however, evince a
 resonance, just like $\epsa$ and $\xia$ and at the same frequency.

 \begin{figure}
\begin{center}
 \begin{tabular}{cc}
\includegraphics[height=8.5cm]{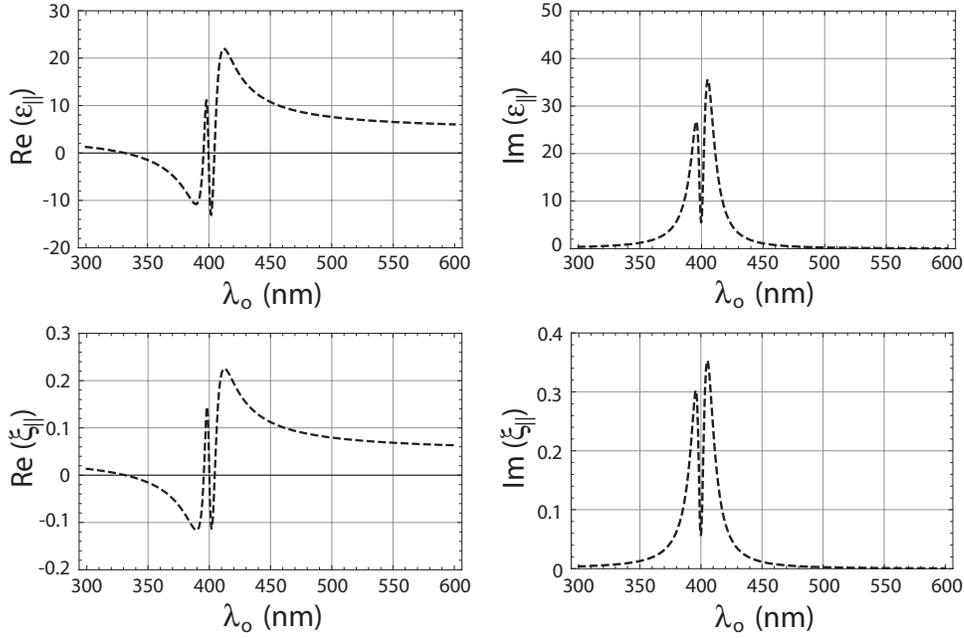}
\end{tabular}
 \end{center}
\caption{ \label{Fig4}   Real and   imaginary parts of $\epsa$,
and  real and  imaginary parts of $\xia$, as  functions of free-space wavelength,
when $f_1=0.58$, $C_2=10^{-3}$, $\lambda_2=400$~nm, $\gamma_2=10^{13}$~rad~s$^{-1}$, and   $\xi_2=0.05$. }
\end{figure}

 \begin{figure}
\begin{center}
 \begin{tabular}{cc}
\includegraphics[height=4cm]{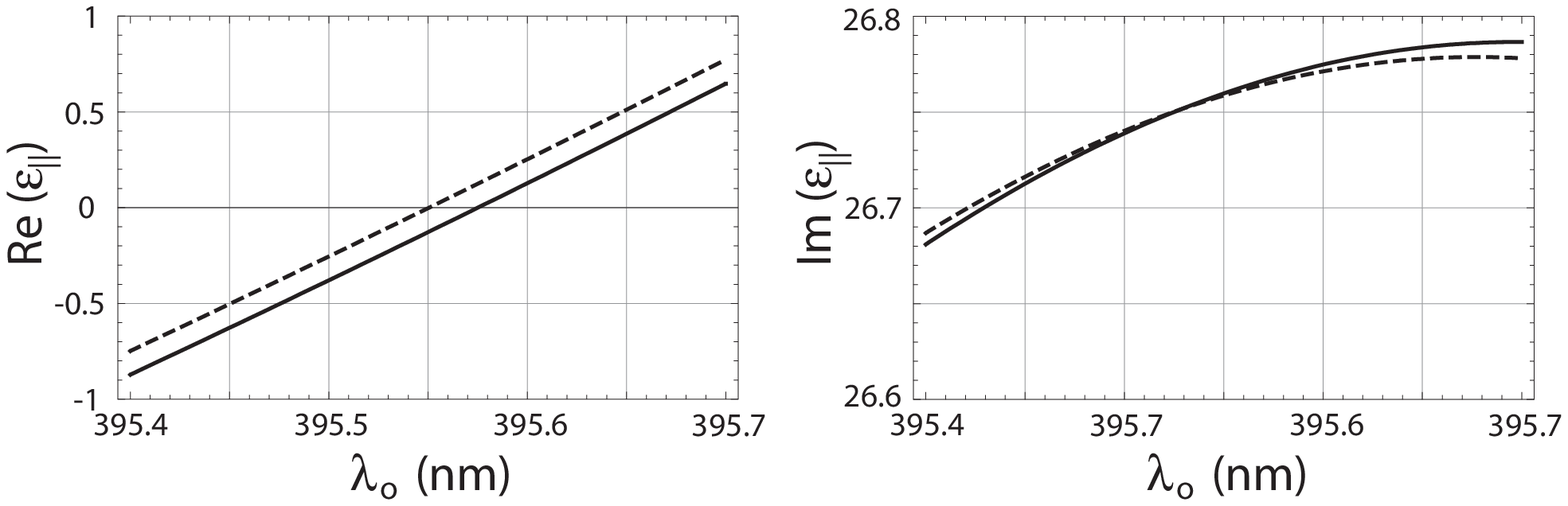}
\end{tabular}
 \end{center}
\caption{ \label{Fig5}   Real and   imaginary parts of $\epsa$ as  functions of free-space wavelength,
when $f_1=0.58$, $C_2=10^{-3}$, $\lambda_2=400$~nm, and $\gamma_2=10^{13}$~rad~s$^{-1}$; $\xi_2=0$ (solid lines) and $\xi_2=0.05$ (dashed lines).}
\end{figure}

Let us now move on to the situation where the relative permittivity of material $2$ is both dissipative and dispersive.
In order to investigate the interplay of structural resonance (evident in Fig.~\ref{Fig2}) and intrinsic
material resonance, we set $\lambda_2=400$~nm; furthermore, $C_2=10^{-3}$ and $\gamma_2=10^{13}$~rad~s$^{-1}$.
Figure~\ref{Fig4} clearly shows the doubly-resonant characteristics of $\epsa$ and $\xia$, arising from this interplay. 
Chirality enhancement is again evident in this figure, as also the enhancement of ${\rm Im}(\epsa)$.

The double resonance in the spectra implies that there are three zero-crossings of ${\rm Re}(\epsa)$. Figure~\ref{Fig5}
presents a highly magnified view of $\epsa$ near one of the three zero-crossings  for both $\xi_2=0$ and $\xi_2=0.05$.
A blueshift of {$0.25$~\AA} is evident, which, we reiterate, is measurable~\cite{OOHR4000}.

We also calculated the linear remittances (i.e., the reflectances and transmittances) of a 20-period multilayered stack that was about a wavelength thick.
The presence of isotropic chirality resulted in the aforementioned blueshifts in the effective constitutive parameters
of the stack, and manifested as small changes in the remittances at oblique incidence. Cross-polarized 
remittances on the order of 1\% were seen when $\xi_2\ne 0$, but were of course  absent for $\xi_2=0$. At normal incidence,
$\xi_2$ played no role at all, which becomes clear from setting $\kappa=0$ on the right side of Eq.~(\ref{P-ubm}); instead,
the stack behaves as a dilute metal.

Our focus on spectroscopic shifts at the boundary of the visible and the ultraviolet regimes is in
accord with the fact that the chirality of many isotropic materials is evident in both regimes.
Examples include suspensions of poly-L-glutamic acid \cite{UK}, solutions of glucose \cite{LCS} and several derivatives of cellulose
\cite{RFMS}, and chiral barbituric acid \cite{YR}. Flooding a multilayered stack containing void regions alternating with
metallic layers with isotropic chiral fluids is a way to dynamically blueshift the zero-crossing of ${\rm Re}\left(\epsa\right)$.

Let us note that we have used $\xi_2=0.05$ in order to illustrate the spectral shifts due to chirality. Chirality paramaters
of natural as well as synthetic materials available in the published literature are smaller. However, values of $\vert\xi_2\vert\sim 0.1$ at microwave frequencies
have recently been obtained \cite{Gomez}. When such large chirality parameters would become
available at optical frequencies, even larger spectral shifts would be possible. Effectively, the shift of the
resonance frequency can enable the use of such stacks as tunable high-pass filters for spectroscopy. The
use of two such filters in tandem can enable a narrow-bandwidth source at almost any desired frequency. Dissipation is, however, a cause for deep concern. We are investigating the possibility of making infrared filters using layers of low-dissipation polaritonic crystals with negative real permittivity (such as LiTaO$_3$ resonance at about 26.7 THz or silicon carbide in the mid-infrared regime) and
isotropic chiral materials. Such filters could be effectively used for spectroscopy of molecular vibro-rotational levels. 

\section{Concluding Remarks}\label{conc}

To conclude, we derived the effective constitutive parameters of a period multilayered stack
whose unit cell comprises a metallic layer and an isotropic chiral layer. We found that 
the stack behaves as an anisotropic
metamaterial with resonant response that is determined primarily  by the volume fraction of
 the metal. The presence of isotropic chirality
can result in small shifts of the resonance frequency for reasonably large
values of the chirality parameter; hence, the sign of one of the two effective
permittivities of the stack can  be switched. The  spectral shifts in the
dielectric properties can be potentially useful for spectroscopic
purposes.

\vspace{1 cm}

\noindent{\bf References}\\


\begin{thebibliography}{99}

\bibitem{Rama-rpp}
Ramakrishna S A 2005
\textit{Rept. Prog. Phys.} \textbf{68} 449

\bibitem{Wood}
Wood B 2007
\textit{Laser Photon. Rev.} \textbf{1} 249


\bibitem{obrien_PRB}  O'Brien S,  McPeake D,  Ramakrishna S A and 
Pendry J B 2004 \textit{Phys. Rev. B} \textbf{ 69} 241101


\bibitem{padilla_Nature2008} Chen H-T,
O'Hara J F, Azad A K., Taylor A J, Averitt R D, Shrekenhamer D B
and Padilla~W~J 2008 
{\it Nature Photon.} {\bf 2} 295

\bibitem{Lspie}
Lakhtakia A 1990
\textit{Selected Papers on Natural Optical Activity} (Bellingham, WA, USA: SPIE)

\bibitem{Lemc}
Lakhtakia A, Varadan V V and Varadan V K 1986
\textit{IEEE Trans. Electromag. Compat.}  \textbf{28} 90 

\bibitem{ChiNih}
Tretyakov S, Nefedov I, Sihvola A, Maslovski S and Simovski C 2003  
\textit{J. Electromag. Waves Appl.} \textbf{17} 695


\bibitem{JPen}
Pendry J B 2004
\textit{Science} \textbf{306} 1353 

\bibitem{LMpre}
Mackay T G and Lakhtakia A 2004
\textit{Phys. Rev. E} \textbf{69} 026602

\bibitem{Mackay-chi}
Mackay T G 2005
\textit{Microw. Opt. Technol. Lett.} \textbf{45} 120;
corrections: 2006 \textbf{47} 406

\bibitem{Bohren}
Bohren C F 2003
in: Weiglhofer W S and Lakhtakia A 2003
\textit{Introduction to Complex Mediums for Optics and Electromagnetics} (Bellingham, WA, USA: SPIE)

\bibitem{VRV}
Varadan V V, Ro R and Varadan V K 1994
\textit{Radio Sci.} \textbf{29} 9

\bibitem{CW}
Chung C Y and Whites K W 1996
\textit{J. Electromag. Waves Appl.} \textbf{10} 1363

\bibitem{Gomez}
G\'omez A, Lakhtakia A, Margineda J,  Molina-Cuberos G J,
N\'u\~{n}ez  J, Ipi\~{n}a J S, Vegas A and Solano M A 2008 
\textit{IEEE Trans. Microw. Theory Tech.} (accepted for publication)


\bibitem{OOHR4000}
http://www.oceanoptics.com/Products/hr4000.asp (24 July 2008)

\bibitem{Weigl94}
Weiglhofer W S 1994
\textit{Int. J. Electron.} \textbf{77} 105

\bibitem{Chen}
Chen H C 1992
\textit{Theory of Electromagnetic Waves} (Fairfax, VA, USA: TechBooks)


\bibitem{L-Brew}
Lakhtakia A 1992
\textit{Optik} \textbf{90} 184

\bibitem{LK}
Lakhtakia A and Krowne C M 2003
\textit{Optik} \textbf{114} 305

\bibitem{LVV-JMR}
Lakhtakia A, Varadan V K and Varadan V V 1992
\textit{J. Mater. Res.} \textbf{8} 917

\bibitem{HN2007}
Hao F and Nordlander P 2007
\textit{Chem. Phys. Lett.} \textbf{446} 115

\bibitem{Pen-PL}
Pendry J B 2000
\textit{Phys. Rev. Lett.} \textbf{85} 3966

\bibitem{SHWERL}
Stahrenberg K,  Herrmann Th, Wilmers K, Esser N, Richter W and Lee M J G
2001 \textit{Phys. Rev. B} \textbf{64} 115111

\bibitem{WL98}
Weiglhofer W S and Lakhtakia A 1998
\textit{Microw. Opt. Technol. Lett.} \textbf{17} 405

\bibitem{UK}
Urry D W and Krivacic J 1970
\textit{Proc. Nat. Acad. Sci. USA} \textbf{65} 845

\bibitem{LCS}
Lin J-Y, Chen K-H and Su D-C 2004
\textit{Opt. Commun.} \textbf{238} 113

\bibitem{RFMS}
Rakhmanberdyev G R, Fedyakova N A, Myagkova N V and Sidikov A 1996
\textit{Chem. Natural Comp.} \textbf{32} 734

\bibitem{YR}
Yeh C and Richardson F S 1975
\textit{Theor. Chim. Acta} {\textbf 39} 197



\end{thebibliography}
\end{document}